\documentclass[nofootinbib,amsfonts,prd,aps]{revtex4}
\usepackage{graphicx}
\begin{document}

\title{An introduction to spherically symmetric 
loop quantum gravity black holes}

\author{Rodolfo Gambini$^{1}$,
Jorge Pullin$^{2}$}
\affiliation {
1. Instituto de F\'{\i}sica, Facultad de Ciencias, 
Igu\'a 4225, esq. Mataojo, 11400 Montevideo, Uruguay. \\
2. Department of Physics and Astronomy, Louisiana State University,
Baton Rouge, LA 70803-4001}

\begin{abstract}
We review recent developments in the treatment of spherically
symmetric black holes in loop quantum gravity. In particular, we
discuss an exact solution to the quantum constraints that represents a
black hole and is free of singularities. We show that new observables
that are not present in the classical theory arise in the quantum
theory. We also discuss Hawking radiation by considering the
quantization of a scalar field on the quantum spacetime. 
\end{abstract}

\maketitle
\section{Introduction}
Spherically symmetric minisuperspaces are an interesting arena to test
ideas of quantum gravity. Classically the system's dynamics is very
rich, including the possibility of gravitational collapse and the
formation of black holes and the critical phenomena discovered by
Choptuik. Quantum mechanically they include the possibility of black
hole evaporation and the issue of information loss and the recently
discussed subject of firewalls. 

Spherically symmetric vacuum space-times were first quantized by
Kastrup and Thiemann \cite{kath} using the original (complex) version
of Ashtekar's variables. Through a series of variable changes and
gauge fixings they were able to complete the quantization. The system
is reduced to one degree of freedom, the mass at infinity, that does
not evolve. Quantum mechanically on has wavefunctions that are
functions of the mass. The resulting quantum states represent
superpositions of black holes with different masses. A year later
Kucha\v{r} \cite{ku} completed the quantization using traditional
metric variables and essentially reached the same results. More
recently, Campiglia et al. \cite{campiglia} completed the quantization
of the exterior of a black hole using the modern version of the
Ashtekar variables and obtained similar results.  In all these
treatments the singularity present in classical black holes was not
eliminated by the quantum treatment. Other groups \cite{interior}
pursued a different strategy for quantization. Using the fact that the
interior of a Schwarzschild black hole is isometric to a
Kantowski--Sachs cosmology, they quantized the system using the
techniques of loop quantum cosmology. As is typically the case in loop
quantum cosmology, the singularity was resolved. There was therefore a
tension between these different treatments. It appeared as in the
first set of treatments one had reduced the theory too much before
quantizing for the techniques of loop quantum gravity to be able to do
much about the singularity. Very recently we \cite{sphericalprl}
completed the quantization of the complete space-time using modern
loop quantum gravity techniques. We found that the singularity is
resolved and remarkably, one could find in closed form the physical
space of states. More recently, we used the resulting quantum
space-time to study a scalar field living on it and compute Hawking
radiation for the quantum black hole. This paper presents a brief
summary of the latter two sets of results. 

\section{Spherically symmetric canonical gravity}

The framework for studying spherically symmetric space-time with the
modern version of the Ashtekar variables was laid down by Bojowald and
Swiderski \cite{boswi}. Choosing triads and connections adapted to
spherical symmetry and after a few changes of variables it is observed
that the Gauss law constraint is eliminated and one is left with a
diffeomorphism constraint along the radial direction and a Hamiltonian
constraint. There are two pairs of canonical variables, one associated
to the radial direction and one to the transverse direction. The
diffeomorphism and Hamiltonian constraints satisfy the same algebra as
in the general non-spherically symmetric case. In particular, they do
not form a Lie algebra. This presents problems at the time of
attempting a Dirac quantization. We \cite{sphericalprl} noted that one
can do a rescaling and combination of the constraints that yield a
Hamiltonian constraint that has an Abelian algebra with itself and the
usual algebra with the diffeomorphism and therefore the total system
of constraints does not contain structure functions in its algebra:
it is a Lie algebra. This opened the door for a Dirac quantization. 

The kinematical quantum arena of loop quantum gravity is given by
states based on spin networks. These are graphs with ``colors'' in
their edges corresponding to representations of $SU(2)$. In each edge
there is an open holonomy constructed with the Ashtekar connection in
a given representation of $SU(2)$ and at the vertices joining edges
the holonomies are contracted with $SU(2)$ intertwiners. There is a
natural diffeomorphism invariant innner product on this space which
essentially states that two spin network states are orthogonal unless
their graphs are equivalent under diffeomorphisms to each other and
the colors on the corresponding edges are the same.  On this space of
states the holonomy of the Ashtekar connection and the (smeared)
triads have well defined actions. An important observation is that
although the holonomy is a well defined operator, the connection is
not. This requires rewriting the equations of the theory in terms of
holonomies and leads to quantum behaviors that may differ from the
ones one sees in the classical theory. This is at the center of the
resolution of the singularities observed, for instance, in loop
quantum cosmology. 

Bojowald and Swiderski \cite{boswi} constructed the kinematical
quantum states for spherically symmetric gravity. The states are one
dimensional spin networks. They are constituted by a ``graph'' which
is a collection of one dimensional edges joined by vertices and each
edge carries a color, labeled by an integer. At the vertices there is
an additional real variable associated with the ``transverse''
canonical pair. As in the full case, the holonomies and smeared triads
are well defined operators. In terms of them one can promote the
Hamiltonian constraint to a well defined quantum operator.

\section{The quantum black hole}
 Remarkably,
considering superpositions of spin networks one can relatively
straightforwardly show what kind of combinations solve the Hamiltonian
constraint exactly. The resulting states depend on a graph, the mass
of the space-time (which is a Dirac observable) and a vector
constituted by the colors of the edges in the spin network. These
states are annihilated by the Hamiltonian constraint but not by the
diffeomorphism constraint. But one can construct states that solve all
the constraints using the group averaging procedure. Essentially one
takes the above states and superposes them with all the states related
by diffeomorhisms. The resulting superposition is invariant under
diffeomorphisms. 

The fact that one can solve the constraints with states with a well
defined number of vertices and a given vector of colors implies that
associated with them are Dirac observables. The total number of
vertices and the vector of colors are Dirac observables. Notice that
these observables do not have any simple classical counterpart. In the
classical theory the only Dirac observable was the mass at infinity,
which is also a Dirac observable at the quantum level.

In gauge theories it is often of interest to study certain gauge
dependent variables evaluated in a given gauge. Gauge dependent
variables in a well defined gauge are gauge invariant information. For
instance one may be interested in studying the metric of the spacetime
associated with the quantum state we discussed above, in a particular
coordinate system. However, one cannot write the metric as an operator
acting on the space of physical states, as it does not commute with
the constraints and therefore as an operator the metric would map us
out of the space of physical states. A technique for capturing the
gauge invariant information present in a gauge dependent variable
evaluated in a particular gauge is the one that Rovelli calls
``evolving constants of motion''. These are Dirac observables
dependent on a (functional) parameter. Different choices of the
parameter correspond to different choices of gauge. It turns out that
one can write the metric of the spacetime as an evolving constant of
motion. In this case the functional parameters specify a system of
coordinates. The gauge invariant information present in it is carried
by the vector of colors of the spin network and the mass of the
space-time. The freedom in choosing coordinates allows to discuss a
complete treatment of the space-time, that is, one can choose the
parameters in such a way that the resulting metric does not have
coordinate singularities at the horizon. As a quantum operator, the
evolving constant of the motion associated with the metric, is
distributional in space-time. It is concentrated at the vertices of
the spin network. One therefore is approximating a classical geometry
via a set of distributions. If one demands that the
evolving constant of the motion be self adjoint, one notices that one
needs to restrict the vector of values of the colors of the spin
network. The value zero and perhaps some other small range of values
has to be excluded. This truncation of the Hilbert space is
consistent: the Hamiltonian and diffeomorphism constraints do not
connect the excluded sector with the remaining one. But the truncation
has a profound consequence: it implies that the region of space-time
where the classical singularity would have been present is excluded. 
Since the geometry was distributional to begin with, excluding a small
region does not cause a problem. The geometry can be extended
through the region where the singularity used to be into another
region of space-time, isometric to the exterior. In that region is a
Cauchy horizon, but since it is isometric to the exterior horizon, it
is likely to be stable. The region is to the future of the space-time,
so this cannot be considered a wormhole. However, the presence of such
a region could have significant implications. For instance, it could
mean that the missing information present in black hole evaporation
went into the new region, if the structure we found here survived
black hole evaporation. 

\section{Black hole evaporation}

Hawking radiation arises because in a curved space-time there is not a
unique definition of the vacuum of a quantum field. The construction
of a vacuum is coordinate dependent. This corresponds to different
observers seeing different number of particles according to their
state of acceleration, a phenomenon already present in flat
space-time. Several vacua have been discussed in the literature
associated with black holes. The Boulware vacuum is obtained using a
coordinate system that covers only the exterior of the black hole. The
modes that are used to construct this vacuum get infinitely blue
shifted as they approach the horizon, as in these coordinates the
modes do not penetrate the horizon. This leads to the development of
singularities in the vicinity of the (past and future) horizon in
physical quantities computed in this vacuum, as the expectation value
of the stress tensor. The Hartle--Hawking vacuum is obtained in
coordinates that cover the whole space-time. It includes ingoing and
outgoing modes, so it does not represent an evaporating black hole but
rather an evaporating black hole with incoming radiation. It does not
lead to the type of singularities seen in the Boulware vacuum. The
Unruh vacuum is also derived in a set of coordinates that cover the
exterior. The Cauchy slices are such that into the past they get
deformed asyptotically into the union of the past horizon and the
past null infinity. This vacuum has singularity issues at the past horizon
as well. Hawking radiation can be computed comparing quantities like
the number operator in the Boulware and Unruh vacua. As is well known
one ends up with a purely thermal spectrum with a temperture inversely
proportional to the mass of the black hole. This implies that the
black hole becomes hotter with time and eventually would appear to
radiate all its mass. Of course this cannot be analyzed in detail
because the calculations of Hawking radiation are done assuming a
fixed background, a hypothesis that does not hold when the black hole
is shrinking rapidly. This has led to decades of discussion of what
exactly is the fate of the black hole. In particular, given that
Hawking's radiation is purely thermal, one can ask what happened to
all the information that went into the creation of the black
hole. Also one could consider forming a black hole by collapsing a
pure quantum state. Since at the end one would allegedly be only left
with thermal radiation, which is in a mixed state, it appears
unitarity is being violated by the process. These issues have led to
the recent proposal of replacing the smooth horizon of black holes by
a ``firewall'' that separates the interior and the extrior
\cite{amps}. Having a quantum theory of gravity raises the hopes that
these issues could be eventually clarified.

In a recent piece of work \cite{hawking} we studied a quantum scalar
field theory on the background given by the quantum black hole. The
approach was to consider the scalar field on a fixed background and
ignore the back reaction. The construction is based on considering
quantum states that are a direct product of the quantum states of
vacuum gravity we discussed above and the states of matter. One takes
the Hamiltonian constraint and evaluates its expectation value on the
gravitational states. The gravitational part of the Hamiltonian
constraint vanishes and the matter part of the Hamiltonian constraint
becomes an operator acting on the matter variables. For this
construction it is key to write the matter part of the Hamiltonian as
an evolving constant of the motion of the vacuum theory, otherwise it
would not be well defined as a quantum operators acting on the vacuum
states. What remains of the matter part of the Hamiltonian after
taking the expectation values is the quantum field theory on a quantum
space time we wish to study. The main effect of the quantum space time
on the quantum field theory is due to the discrete nature of the
quantum geometry. What would have been the partial differential
equations of a scalar field on a curved space time now become
difference equations. 

The discreteness has important implications for the quantum scalar
field. As expected, the quantum geometry acts as a natural regulator
of the matter theory. It implies the existance of a maximum frequency
for the quantum modes. One can choose the level of
discreteness present by choosing the state of vacuum gravity used to
take the expectation value of the Hamiltonian constraint. Naturally,
in order to approximate well a smooth space-time, one would like to
choose a small spacing between the vertices of the spin network. One
would also like to choose the vector of colors of the spin network in
such a way that adjacent colors do not differ by much. Since the
colors are associated with the eigenvalues of the metric and these
with the values of the areas of the surfaces of symmetry, if one
choose colors that differ abruptly there would be jumps in the
geometry. The quantization of the area in loop quantum gravity in turn
imposes a lower bound on how close the vertices of the spin network
can be, it leads to a limit of $\ell_{\rm Planck}/(2r)$. Notice that
this number is, for macroscopic black holes, considerably smaller than
Planck's length, so this implies the resulting discrete quantum theory
provides an excellent approximation to the continuum theory. So most
results concerning black hole evaporation hold as a
consequence. However, the presence of discreteness brings an important
difference. Since there is a maximum frequency, the problems that
originated the singularities in the Boulware and Unruh vacuum are
absent. 

The discrete quantum theory naturally implements the cutoff in the
propagators discuss by \cite{valencia} and lead to the same formula for
the Hawking radiation discussed in that reference. At least for
macroscopic black holes and for typical frequencies, it implies a very
small correction to the formula of Hawking's temperature. 

\section{Summary}

Spherically symmetric vacuum space times can be quantized in loop
quantum gravity. The physical space of states can be found in closed
form. The metric can be realized as a parameterized Dirac
observable. As a function of space time it is distributional, taking
values only at the vertices of the one-dimensional spin network. For
the metric to be self adjoint, the region where classically one would
have a singularity is absent. Space-time can be continued to another
region isometric to the region exterior to where the singularity used
to be. The presence of the new region could have implications for the
black hole information paradox.

On the quantum space time one can study a quantum scalar field
theory. The main effect of the quantum background space-time is to
discretize the equations of the field theory due to the discreteness
of space in loop quantum gravity. The discrete equations approximate
ordinary quantum field theory very well and most usual results
hold. However, the presence of a maximum frequency due to the
discreteness eliminates the singularities associated with some of the
quantum vacua. Hawking radiation can be computed and one ends up with
a formula that differs little from the ordinary for macroscopic black
holes and typical frequencies.

We wish to thank Iv\'an Agull\'o and Abhay Ashtekar for discussions.
This work was supported in part by grant NSF-PHY-1305000, funds of the
Hearne Institute for Theoretical Physics, CCT-LSU and Pedeciba.

\end{document}